\documentclass[prX,showpacs,twocolumn,nofootinbib]{revtex4}

\usepackage{amssymb}    
\usepackage{amsmath}
\usepackage{slashed}
\newcommand{\be}{\begin{equation}}
\newcommand{\ee}{\end{equation}}
\newcommand{\ba}{\begin{eqnarray}}
\newcommand{\ea}{\end{eqnarray}}

\def\a{\alpha}
\def\b{\beta}
\def\d{\delta}
\def\e{\epsilon}
\def\ve{\varepsilon}
\def\f{\phi}
\def\vf{\varphi}
\def\g{\gamma}
\def\h{\eta}
\def\j{\psi}

\def\l{\lambda}
\def\m{\mu}
\def\n{\nu}

\def\r{\rho}
\def\s{\sigma}

\def\x{\xi}

\def\D{\Delta}

\def\G{\Gamma}

\def\O{\Omega}

\def\S{\Sigma}

\def\cf{{\cal F}}

\def\co{{\cal O}}

\def\cs{{\cal S}}

\newcommand{\ov}{\overline}

\newcommand{\wh}{\widehat}

\newcommand{\aand}{\;\;\;\mbox{and}\;\;\;}
\newcommand{\pa}{\partial}

\def\sl#1{\rlap{\hbox{$\mskip 1 mu /$}}#1}

\def\I{\leavevmode\hbox{\small1\kern-3.8pt\norfeynmpmalsize1}}

\begin{document}
%\draft
%\twocolumn
\title{No radiative generation of Chern-Simons-like term in Lorentz-violating QED: dealing with IR divergences}
\author{O.M. Del Cima}
\email{oswaldo.delcima@ufv.br}
\affiliation{Universidade Federal de Vi\c cosa (UFV),\\
Departamento de F\'\i sica - Campus Universit\'ario,\\
Avenida Peter Henry Rolfs s/n - 36570-900 -
Vi\c cosa - MG - Brazil.}

\author{D.H.T. Franco}
\email{daniel.franco@ufv.br}
\affiliation{Universidade Federal de Vi\c cosa (UFV),\\
Departamento de F\'\i sica - Campus Universit\'ario,\\
Avenida Peter Henry Rolfs s/n - 36570-900 -
Vi\c cosa - MG - Brazil.}

\author{O. Piguet}
\email{opiguet@pq.cnpq.br}
\affiliation{Universidade Federal de Vi\c cosa (UFV),\\
Departamento de F\'\i sica - Campus Universit\'ario,\\
Avenida Peter Henry Rolfs s/n - 36570-900 -
Vi\c cosa - MG - Brazil.}

\date{\today}
%\maketitle
%{In honor of Prof. Manfred Schweda (1939-2017)}
%===================================================================
\begin{abstract}
The issue intensively claimed in the literature on the generation of a CPT-odd and Lorentz violating Chern-Simons-like term by radiative corrections owing to a CPT violating interaction -- the axial coupling of  fermions with a constant vector field $b_\m$ -- is mistaken. The presence of massless gauge field triggers IR divergences that might show up from the UV subtractions, therefore, so as to deal with the (actual physical) IR divergences, the Lowenstein-Zimmermann subtraction scheme, in the framework of BPHZL renormalization method, has to be adopted. The proof on the non generation of such a Chern-Simons-like term is done, independent of any kind of regularization scheme, at all orders in perturbation theory.
\end{abstract}
\maketitle

{\it In honor of Prof. Manfred Schweda (1939-2017)}

%===================================================================
%%%%%%%%%%%%%%%%%%%%%%%%%%%%%%%%%%%%%%%%%%%%%%
\section{Introduction}
%%%%%%%%%%%%%%%%%%%%%%%%%%%%%%%%%%%%%%%%%%%%%%
Perturbative field models with symmetry breaking were investigated from the point of view of 
the theory of renormalization in the pioneering work of Symanzik \cite{Sym1,Sym2} and treated in a way 
that we can consider as definitive, by Becchi-Rouet-Stora \cite{Stora1,Stora2,Stora3}. However, several recent works, dealing in particular with field theories with Lorentz symmetry breaking, do not consider very carefully how the symmetry is broken, not taking into account the requirements that Symanzik-Becchi-Rouet-Stora have shown to be necessary. In this article, all our analysis will be based on a general iterative scheme called Algebraic Renormalization\footnote{It should be emphasized that, based on the method suggested by the Epstein-Glaser construction \cite{Epstein-Glaser,Zimmermann}, the algebraic method of renormalization was {\it seeded} by Stora in Ref.\cite{Stora4}.} \cite{piguet-rouet,PigSor,Man}. In the algebraic approach, in order to study the renormalizability of models characterized by a system of Ward identities, without referring to any special regularization procedure, two steps must be followed. In the first step, for a power-counting renormalizable model, at the level of the radiative corrections, one investigates the preservation of the symmetries, or the determination of all possible anomalies. This amounts to find the solution of the cohomology of its symmetry group: trivial elements (co-boundaries) correspond to breakings which can be compensated by non-invariant counterterms, whereas the non-trivial elements are the possible anomalies. These cohomology conditions are a generalization of  the Wess-Zumino consistency condition \cite{Wess-Zumino} used in order to compute the possible anomalies of the Ward identities in Yang-Mills theories. In a second step, we check the stability of the classical action -- which ensures that the quantum corrections do not produce counterterms corresponding to the renormalization of parameters not already present in the classical model.

Let us emphasize that the algebraic renormalization scheme is based on a set of general theorems of renormalization theory, collected under the name of Quantum Action Principle (QAP) \cite{Lam,Lowenstein,Brenneke-Dutsch}. These theorems deal with the whole of Feynman graphs combinatorics and integrability, so that explicit graph considerations are unnecessary -- unless one looks for explicit quantitative results for applications to physics, of course. As Stora said:``{\it Use the theorems!}\,''.

The quantum electrodynamics (QED) \cite{ext-QED} with violation of Lorentz and CPT \cite{greenberg-chaichian}  have been studied intensively. Among several issues, the possible generation of a Chern-Simons-like term induced by radiative corrections arising from a CPT and Lorentz violating term in the fermionic sector has been a recurrent theme in the literature. We particularly mention the following works \cite{Ref1,Ref2,Ref3,Ref4,Ref5,Ref7,Ref8,Ref9,Ref10,Ref11,Ref11',Ref12,Ref12',Ref12'',Ref13,Ref14,Brito,Ref15,Nos1,Nos2,Mariz} 
(and references cited therein), where many controversies have emerged from the discussion whether this
Chern-Simons-like term could be generated by means of radiative corrections arising from the axial coupling of charged fermions to a constant vector $b_\mu$ responsible for the breakdown of Lorentz symmetry.

In this work, we reassess the discussion on the radiative generation of a Chern-Simons-like term induced from quantum corrections in the extended QED. We show, to all orders in perturbation theory, that a CPT-odd and Lorentz violating Chern-Simons-like term, {\it definitively}, is not radiatively induced by the axial coupling of the fermions with the constant vector $b_\mu$. The proof of this fact is based on general theorems of perturbative quantum field theory (see \cite{piguet-rouet,PigSor,Man} and references therein), where the Lowenstein-Zimmermann subtraction scheme in the framework of Bogoliubov-Parasiuk-Hepp-Zimmermann-Lowenstein 
(BPHZL) renormalization method \cite{Low} is adopted. The former has to be introduced, owing to the presence 
of massless gauge field, so as to subtract infrared (IR) divergences that should arise from the ultraviolet (UV) subtractions.

This article is structured as follows: in Section \ref{sec2}, the quantum electrodynamics (QED) with a term which violates Lorentz and CPT (extended QED) is introduced, it is established all continuous and discrete simmetries at the classical level, as well as determined the ultraviolet and infrared dimensions of all the fields; the behaviour of the extended QED at the quantum level is analyzed in Section \ref{sec3}; Section \ref{conc} are left to the final comments and conclusions.

%%%%%%%%%%%%%%%%%%%%%%%%%%%%%%%%%%%%%%%%%%%%%%
\section{The model at the classical level}
\label{sec2}
%%%%%%%%%%%%%%%%%%%%%%%%%%%%%%%%%%%%%%%%%%%%%%
We start by considering an action for extended QED with a term which violates the
Lorentz and CPT symmetries in the matter sector only. In the tree approximation,
the classical action of extended QED with one Dirac spinor that we are considering
here is given by:
\be
\Sigma^{(s-1)}=\Sigma_{\rm S}+\Sigma_{\rm SB}+\Sigma_{\rm IR}
+\Sigma_{\rm gf}+\Sigma_{\rm ext}~,\label{1}
\ee
where
\be
\Sigma_{\rm S}=
\int d^{4}x~\left\{-\frac{1}{4}F^{\mu\nu}F_{\mu\nu} + i\bar{\psi}\gamma^{\mu}D_{\mu}\psi -
m \bar{\psi}\psi\right\}~,
\ee
is the symmetric part of $\Sigma^{(s-1)}$ under gauge and Lorentz transformations, and $D_\m\psi\equiv(\partial_{\mu}+ieA_{\mu})\psi$. The term
\be
\Sigma_{\rm SB}=-\int d^{4}x~b_{\mu}\bar{\psi}\gamma_{5}\gamma^{\mu}\psi~,\label{SB}
\ee
is the non symmetric part of $\Sigma^{(s-1)}$. It violates CPT symmetry and breaks the manifest Lorentz
covariance on account of a constant vector $b_{\mu}$, which selects a preferential direction in Minkowski space-time, breaking its isotropy. In addition to,
\be
\Sigma_{\rm IR}=\int d^{4}x~\frac{1}{2}\mu^2 (s-1) A_\mu A^\mu~,
\ee
is the Lowenstein-Zim\-mer\-mann mass term for the (massless) photon field. A Lowenstein-Zim\-mer\-mann mass term must be introduced in order to enable subtractions in momenta space without introducing spurious infrared (IR) singularities. The Lowenstein-Zimmermann parameter $s$ lies in the interval 
$0 \leq s \leq 1$ and plays the role of an additional subtraction variable (as the external momentum) in the 
BPHZL renormalization program, such that the theory describing a really massless particle is
recovered for $s=1$. It should be comment that the Lowenstein-Zimmermann mass term for the photon 
field does not spoil gauge invariance at the quantum level\footnote{This was investigated in details for the QED in Ref.\cite{low-schroer} using the BPHZ scheme.}; this is a peculiarity of the abelian case \cite{piguet-rouet}. Finally, in order to quantize the model, a gauge-fixing 
\be
\Sigma_{\rm gf}=\int d^{4}x~\left\{b \partial_\mu A^\mu + \frac{\xi}{2}b^2 + \overline{c} {\square}c \right\}~,
\label{gf}
\ee
is added, together with the term, $\Sigma_{\rm ext}$, by coupling the non-linear Becchi-Rouet-Stora 
(BRS) transformations to external sources
\be
\Sigma_{\rm ext}=\int d^{4}x~\left\{\overline{\Omega}{\mathfrak s}
\psi-{\mathfrak s}{\overline \psi}\Omega\right\}~.
\label{ext}
\ee

%%%%%%%%%%%%%%%%%%%%%%%%%%%%%%%%%%%%%%%%%%%%%%
\subsection{Continuous symmetries:}
%%%%%%%%%%%%%%%%%%%%%%%%%%%%%%%%%%%%%%%%%%%%%%
The BRS transformations are given by:
\ba
&&{\mathfrak s} \psi=ic \psi~,~~{\mathfrak s} \overline{\psi}=-ic \overline \psi~;\nonumber \\
&&{\mathfrak s} A_\mu=-\frac{1}{e} \partial_\mu c~,~~ {\mathfrak s} c=0~; \label{BRS} \\
&&{\mathfrak s} {\overline c}=\frac{1}{e}b~~,~~{\mathfrak s} b=0~; \nonumber
\ea
where $c$ is the ghost field, ${\overline c}$ is the antighost field and $b$ is the Lautrup-Nakanishi 
field \cite{lautrup-nakanishi}, respectively. Although not massive, the Faddeev-Popov ghosts, $c$ and ${\overline c}$, are free fields, they decouple, therefore, there is no need to introduce a Lowenstein-Zimmermann mass term for them.

The BRS invariance of the action is expressed in a functional way by the Slavnov-Taylor identity
\be
\cs(\S^{(s-1)})=0~,
\label{slavnovident}
\ee
where the Slavnov-Taylor operator $\cs$ is defined, acting on an arbitrary functional $\cf$, by
\ba
\cs(\cf)&\!\!=\!\!&\int{d^4 x}~\biggl\{-{1\over e}{\pa}^\mu c {\d\cf\over\d A^\mu} + 
{1\over e}b {\d\cf\over\d {\ov c}} + \nonumber\\
&&+{\d\cf\over\d \ov\O}{\d\cf\over\d \j} - 
{\d\cf\over\d \O}{\d\cf\over\d \ov\j}\biggl\}~.\label{slavnov}
\ea
The corresponding linearized Slavnov-Taylor operator reads
\ba
\cs_\cf&\!\!=\!\!&\int{d^4 x}~\biggl\{-{1\over e}{\pa}^\mu c {\d\over\d A^\mu} + 
{1\over e}b {\d\over\d {\ov c}} + {\d\cf\over\d \ov\O}{\d\over\d \j} + 
{\d\cf\over\d \j}{\d\over\d \ov\O} + \nonumber \\
&&-{\d\cf\over\d \O}{\d\over\d \ov\j} - {\d\cf\over\d\ov\j}{\d\over\d \O} \biggl\}~.\label{slavnovlin}
\ea
The following nilpotency identities hold:
\ba
&&\cs_\cf\cs(\cf)=0~,~~\forall\cf~, \label{nilpot1} \\
&&\cs_\cf\cs_\cf=0~~{\mbox{if}}~~\cs(\cf)=0~. \label{nilpot3}
\ea
In particular, $(\cs_\S^{(s-1)})^2=0$, since the action $\S^{(s-1)}$ obeys
the Slavnov-Taylor identity (\ref{slavnovident}). The operation of
$\cs_{\S^{(s-1)}}$ (\ref{slavnovlin}) upon the fields and the external sources reads
\ba
&&\cs_{\S^{(s-1)}}\f=s\f~,~~\f=\{\j,\ov\j,A_\m,c,{\ov c},b\}~, \\
&&\cs_{\S^{(s-1)}}\O=-{\d\S^{(s-1)}\over\d\ov\j}~, \\ 
&&\cs_{\S^{(s-1)}}\ov\O_+={\d\S^{(s-1)}\over\d\j}~. 
\ea

In addition to the Slavnov-Taylor identity (\ref{slavnovident}), the classical action 
$\S^{(s-1)}$ (\ref{1}) is characterized by the gauge condition, the ghost equation and the
antighost equation:
\ba
&&{\d\S^{(s-1)}\over\d b}=\pa^\m A_\m + \x b~,\label{gaugecond} \\
&&{\d\S^{(s-1)}\over\d \ov c}=\square c~,\label{ghostcond} \\
&&-i{\d\S^{(s-1)}\over\d c}=i\square{\ov c} + \ov\O\j - \ov\j\O~.\label{antighostcond}
\ea

The action $\S^{(s-1)}$ (\ref{1}) is invariant also with respect to the rigid symmetry
\be
W_{\rm rigid} \S^{(s-1)}=0~, \label{rigidcond}
\ee
where the Ward operator, $W_{\rm rigid}$, is defined by
\be
W_{\rm rigid}=\int{d^4 x}~\biggl\{\j{\d\over\d \j} - \ov\j{\d\over\d \ov\j} 
+ \O{\d\over\d \O} - \ov\O{\d\over\d \ov\O} \biggr\}~. 
\ee

On the other hand, the Lorentz symmetry is broken by the presence of the 
constant vector $b_\m$. The fields $A_\m$ and $\j$ transform under
infinitesimal Lorentz transformations -- $\d x^\m$ =$\e^\m{}_\n x^\n$ ($\e_{\m\n}=-\e_{\n\m}$) -- 
in such a way that
\ba
\delta_{\rm L} A_\mu &\!\!=\!\!&-\e^\l{}_\n x^\n\pa_\l A_\m + \e_\m{}^\n A_\n
\equiv \frac12 \e^{\a\b} \d_{\rm L\a\b} A_\m~; \\
\delta_{\rm L}\j  &\!\!=\!\!&-\e^\l{}_\n x^\n\pa_\l \j
-\frac{i}{4}\e^{\m\n}\s_{\m\n}\j \equiv \frac12 \e^{\a\b} \d_{\rm L\a\b} \j~, \\
&&{\mbox{where}}~~ \sigma_{\mu\nu}=\frac{i}{2}[\gamma_\mu,\gamma_\nu]~. 
\label{lor-var-A-psi}
\ea 

It has to be pointed out that, since the Lorentz breaking $\Sigma_{\rm SB}$ (\ref{SB}) 
is not linear in the quantum fields, it shall be submitted to renormalization.
Nevertheless, it is a soft breaking, its ultraviolet (UV) power-counting   
dimension is less than 4, namely 3. A model with soft
symmetry breakings is renormalizable if the radiative corrections do not
induce a breakdown of the symmetries by terms of UV 
dimension equal to 4, the hard breakings \cite{Sym1,Sym2}. Bearing in mind  
the Weinberg's theorem \cite{weinberg}, it can be concluded that the symmetry of the 
theory, in the asymptotic deep euclidean region of momentum space, remains preserved 
by radiative corrections. In order to control the Lorentz breaking and, in 
particular, its power-counting properties, by following \cite{Sym1,Sym2}, and \cite{Balasin} 
for the specific case of Lorentz breaking, we introduce
an external field $\b_\m\equiv\b_\m(x)$, with UV and IR dimensions equal to 1, which 
transforms under Lorentz transformations as
\be
\delta_{\rm L}\b_\mu=-\e^\l{}_\n x^\n\pa_\l \b_\m+
\e_\m{}^\n (\b_\n+b_\n)
\equiv \frac12 \e^{\a\b} \d_{\rm L\a\b} \b_\m~.\label{lor-var-beta}
\ee
The Ward operator associated to the Lorentz symmetry reads 
\be 
{W}_{\rm L}=\frac12 \e^{\a\b}{W}_{\rm L\a\b}~,
\ee
where
\be
{W}_{\rm L\a\b}=
\int d^4x\sum_{\vf=A_\m,\j,\ov\j,\b_\m}\d_{\rm L\a\b}\vf\frac{\d}{\d \vf}~.\label{Lor-w-op}
\ee
By adding, to the action $\S^{(s-1)}$ (\ref{1}), a term depending on $\b_\mu$, such as:
\be
\widetilde\S^{(s-1)} = \S^{(s-1)} - \int d^4x~\b_\m\ov\j\g_5\g^\m\j~,\label{25}
\ee
it can be verified the following classical Ward identity
\be
{W}_{\rm L\a\b}\widetilde\Sigma^{(s-1)}=0~,\label{tilde-Lor-Ward}
\ee
so that, at $\b_\m=0$, it reduces to the broken Lorentz Ward identity 
\be
{W}_{\rm L\a\b}\Sigma^{(s-1)} = 
-b_{[\a}\int d^4x~\ov\j\g_5\g_{\b]}\j~.\label{Lor-Ward}
\ee
Owing to the fact that the external field $\beta_\mu$ is coupled (\ref{25}) to the gauge invariant axial current ($j_5^\mu=\ov{\psi}\gamma_{5}\gamma^{\mu}\psi$), it is assumed to be BRS invariant in order to preserve gauge invariance,
\be
{\mathfrak s} \int d^4x\,\b_\m\ov\j\g_5\g^\m\j
=0 ~\Longrightarrow~ {\mathfrak s}\beta_\mu=0~.
\ee
Consequently, the action $\widetilde{\Sigma}^{(s-1)}$ (\ref{25}) satisfies the same Slavnov identity (\ref{slavnovident}) as the action $\Sigma^{(s-1)}$ (\ref{1}): 
\be
\cs(\widetilde{\Sigma}^{(s-1)})=0~, \label{tilde-slavnov}
\ee
together with the following identities:
\ba
&&{\d\widetilde\S^{(s-1)}\over\d b}=\pa^\m A_\m + \x b~,\label{tilde-gaugecond} \\
&&{\d\widetilde\S^{(s-1)}\over\d \ov c}=\square c~,\label{tilde-ghostcond} \\
&&-i{\d\widetilde\S^{(s-1)}\over\d c}=i\square{\ov c} + \ov\O\j - \ov\j\O~, \label{tilde-antighostcond} \\
&&W_{\rm rigid} \widetilde\S^{(s-1)}=0~, \label{tilde-rigidcond}
\ea

%%%%%%%%%%%%%%%%%%%%%%%%%%%%%%%%%%%%%%%%%%
\subsection{Discrete symmetries:}
%%%%%%%%%%%%%%%%%%%%%%%%%%%%%%%%%%%%%%%%%%
\subsubsection*{Charge conjugation:}
Assuming the Dirac representation of the $\gamma$-matrices \cite{itzykson}, the 
charge conjugation transformations read: 
\ba
&&\psi \stackrel{\rm C}{\longrightarrow} C{\ov{\psi}}^T=i\gamma^{2}\gamma^{0}{\ov{\psi}}^T~,\nonumber \\
&&A_\mu \stackrel{\rm C}{\longrightarrow} -A_{\mu}~, \nonumber \\
&&C\gamma_{\mu}C^{-1} = - \gamma_{\mu}^{T}~,\label{9}
\ea
then, it is verified that all terms in the action $\widetilde{\Sigma}^{(s-1)}$ (\ref{25}) are invariant under
charge conjugation.

\subsubsection*{Parity:}
The parity transformations are given by:
\ba
&&x^{\mu} \stackrel{\rm P}{\longrightarrow} x_{\mu}~, \nonumber\\
&&\psi \stackrel{\rm P}{\longrightarrow} P\psi=\gamma^{0}\psi~, \nonumber\\
&&A^{\mu} \stackrel{\rm P}{\longrightarrow} A_{\mu}~, \nonumber \\
&&P\gamma_{\mu}P^{-1} = - (-1)^{\h_{\m\m}} \gamma_{\mu}~,\label{11}
\ea
where in this case all terms of the action $\widetilde{\Sigma}^{(s-1)}$ (\ref{25}), except the Lorentz breaking term $\Sigma_{\rm SB}$ (\ref{SB}), are invariant under parity.

\subsubsection*{Time reversal:} 
The time reversal transformations follow:
\ba
&&x^{\mu} \stackrel{\rm T}{\longrightarrow} -x_{\mu}~, \nonumber\\
&&\psi \stackrel{\rm T}{\longrightarrow} T\psi=i\gamma^{1}\gamma^{3}\psi~,\nonumber \\
&&A^{\mu} \stackrel{\rm T}{\longrightarrow} A_{\mu}~, \nonumber \\
&&T\gamma^{\mu}T^{-1} = \gamma_{\mu}^{T}~,\label{13}
\ea
where it is verified that the Lorentz breaking term $\Sigma_{\rm SB}$ (\ref{SB}) is not invariant under time reversal, whereas the other terms in the action $\widetilde{\Sigma}^{(s-1)}$ (\ref{25}) remain invariant. 

Consequently, the action $\widetilde{\Sigma}^{(s-1)}$ (\ref{25}), has CPT symmetry broken by the Lorentz breaking term, $\Sigma_{\rm SB}$ (\ref{SB}): 
\be
\ov{\psi}b_{\mu}\gamma_{5}\gamma^{\mu}\psi \stackrel{\rm CPT}{\xrightarrow{\hspace*{0,75cm}}} -\ov{\psi}b_{\mu}\gamma_{5}\gamma^{\mu}\psi~.\label{15}
\ee

%%%%%%%%%%%%%%%%%%%%%%%%%%%%%%%%%%%%%%%%%%
\subsection*{UV and IR dimensions:}
%%%%%%%%%%%%%%%%%%%%%%%%%%%%%%%%%%%%%%%%%%
Switching off the coupling constant ($e$) and taking the free part of the action (\ref{1}), 
the tree-level propagators in momenta space, for all the fields, read: 
\ba
&&\Delta_{\ov\psi\psi}(k)=i\frac{\sl{k}+m}{k^2-m^2}~,\label{propk++--} \\
&&\Delta^{\mu\nu}_{AA}(k,s)=-i\biggl\{\frac{1}{k^2-M^2(s-1)^2}\biggl(\eta^{\mu\nu}-\frac{k^\mu k^\nu}{k^2}\biggr) + \nonumber \\
&&~~~~~~~~~~~~~~~~~+ \frac{\xi}{k^2-\xi\,M^2(s-1)^2}\frac{k^\mu k^\nu}{k^2} \biggr\}~, \label{propkAA} \\
&&\Delta^\mu_{Ab}(k)=\frac{k^\mu}{k^2}~,~~\Delta_{bb}(k)=0~, \label{propkAbbb} \\
&&\Delta_{{\overline c}c}(k)=-i\frac{1}{k^2}~.\label{propkcbc}
\ea

The ultraviolet (UV) and infrared (IR) dimensions of any fields, $X$ and $Y$, are given by the UV and IR asymptotic behaviour of their propagator ($\Delta_{XY}(k,s)$), $d_{XY}$ and $r_{XY}$, respectively, defined as follows: 
\ba
&&d_{XY}={\overline{\rm deg}}_{(k,s)}\Delta_{XY}(k,s)~, \\
&&r_{XY}={\underline{\rm deg}}_{(k,s-1)}\Delta_{XY}(k,s)~,
\ea
where the upper degree ${\overline{\rm deg}}_{(k,s)}$ gives the asymptotic power 
for $(k,s)\rightarrow \infty$ whereas the lower degree ${\underline{\rm deg}}_{(k,s-1)}$ 
gives the asymptotic power for $(k,s-1) \rightarrow 0$. The UV ($d$) and IR ($r$) 
dimensions of the fields, $X$ and $Y$, shall respect the following inequalities:
\be
d_X + d_Y \geqslant 4 + d_{XY} \aand  r_X + r_Y \leqslant 4 + r_{XY}~. 
\label{uv-ir}
\ee

In summary, the UV ($d$) and IR ($r$) dimensions -- which are those involved in the Lowenstein-Zimmermann 
subtraction scheme \cite{Low} -- as well as the ghost numbers ($\Phi\Pi$) and the Grassmann parity (GP) of 
all fields are displayed in Table \ref{table1}. It should be stressed that the statistics among the fields is defined as follows: the integer spin fields with odd ghost number, as well as, the half integer spin fields with even ghost number anticommute among themselves. However, the other fields commute with the formers and also among themselves.

%%%%%%%%%%%%%%%%%%%%%%%%%%%%%%%%%%%%%%%%%%%%%%%%%%%%%%%%%%%%%%%%%%%%%%
\begin{table}
\begin{center}
\begin{tabular}{|c||c|c|c|c|c|c|c|c|c|}
\hline
    &$A_\mu$ &$\psi$ &$c$ &${\overline c}$ &$b$ &$\Omega$ &$\beta_\mu$ &$s-1$ &$s$ \\
\hline\hline
$d$ &1 &${3/2}$ &0 &2 &2 &5/2 &1 &1 &1 \\
\hline
$r$ &1 &2 &0 &2 &2 &2 &1 &1 &0 \\
\hline
$\Phi\Pi$&0 &0 &1 &$-1$ &0 &$-1$ &0 &0 &0 \\
\hline
$GP$&0 &1 &1 &1 &0 &0 &0 &0 &0 \\
\hline
\end{tabular}
\end{center}
\caption[]{UV ($d$) and IR ($r$) dimensions, ghost number ($\Phi\Pi$) and Grassmann parity ($GP$).}
\label{table1}
\end{table}
%%%%%%%%%%%%%%%%%%%%%%%%%%%%%%%%%%%%%%%%%%%%%%%%%%%%%%%%%%%%%%%%

%%%%%%%%%%%%%%%%%%%%%%%%%%%%%%%%%%%%%%%%%%
\section{The model at the quantum level}
\label{sec3}
%%%%%%%%%%%%%%%%%%%%%%%%%%%%%%%%%%%%%%%%%%
Following Symanzik -- ``{\it whether you like it or not, you have to include in the lagrangian all counter terms consistent with locality and power-counting, unless otherwise constrained by Ward identities}''  \cite{symanzik} -- we present next, the perturbative quantization of the extended QED model, using the algebraic renormalization method \cite{PigSor,Man}. Our aim is to prove that the full quantum model has the same properties as the classical model, namely, we have to demonstrate that, at the quantum level, the Ward identity related to the Lorentz symmetry (\ref{tilde-Lor-Ward}) and the Slavnov-Taylor identity  associated to the gauge symmetry (\ref{tilde-slavnov}) are satisfied at all orders in perturbation theory:
\ba
&&{W}_{\rm L\a\b}\G^{(s-1)}\Big|_{s=1}=0~,\label{quantum-Lor-Ward}\\
&&\cs({\G}^{(s-1)})\Big|_{s=1}=0~. \label{quantum-slavnov} 
\ea
In order to study the renormalizability of models characterized by a system of Ward 
identities, without referring to any special regularization scheme, two procedures 
must be followed \cite{PigSor,Man}. First, we search for possible
anomalies of the Ward identities through an analysis of the Wess-Zumino
consistency condition. Second, we verify the stability of the classical action, which guarantees  
that the quantum corrections do not produce counterterms corresponding to the renormalization 
of parameters which are not already present at the classical level.

%%%%%%%%%%%%%%%%%%%%%%%%%%%%%%%%%%%%%%%%%%
\subsection{The Lorentz-Ward and the Slavnov-Taylor identities: in search for anomalies}
%%%%%%%%%%%%%%%%%%%%%%%%%%%%%%%%%%%%%%%%%%
At the quantum level the vertex functional, $\G^{(s-1)}$, which coincides with the classical action, 
$\widetilde\S^{(s-1)}$ (\ref{25}), at zeroth order in $\hbar$,
\be
\G^{(s-1)}=\widetilde\S^{(s-1)} + {\co}(\hbar)~,\label{vertex}
\ee
has to satisfy the same constraints as the classical action, namely Eq.(\ref{tilde-Lor-Ward}) and Eqs.(\ref{tilde-slavnov})-(\ref{tilde-rigidcond}).

According to the Quantum Action Principle \cite{Lam,Lowenstein,Brenneke-Dutsch}, due to radiative corrections, the Lorentz symmetry Ward identity (\ref{tilde-Lor-Ward}) and the Slavnov-Taylor identity (\ref{tilde-slavnov}) develop quantum breakings:
\ba
&&{W}_{\rm L\a\b}\G^{(s-1)}\Big|_{s=1}=\D_{\rm L\a\b} \cdot \G^{(s-1)}\Big|_{s=1} \nonumber\\
&&~~~~~~~~~~~~~~~~~~~~~= \D_{\rm L\a\b} + {\co}(\hbar \D_{\rm L\a\b})~,\label{lorentzbreak}\\
&&\cs(\G^{(s-1)})\Big|_{s=1}=\D \cdot \G^{(s-1)}\Big|_{s=1} \nonumber\\
&&~~~~~~~~~~~~~~~~~~= \D_{\rm g} + {\co}(\hbar \D_{\rm g})~,\label{slavnovbreak}
\ea
where $\D_{\rm L\a\b}\equiv\D_{\rm L\a\b}|_{s=1}$ and $\D_{\rm g}\equiv\D_{\rm g}|_{s=1}$ are 
integrated local functionals, taken at $s=1$, with ghost number one and, UV and IR dimensions 
bounded by $\d\le4$ and $\r\ge4$, respectively.

The validity of the Lorentz Ward identity has been proved in \cite{Balasin} by using the Whitehead's lemma 
for semi-simple Lie groups, which states the vanishing of the first cohomology of such kind of group \cite{Stora1,Stora3}. Here\footnote{See details in \cite{Nos1}.}, this means that the Lorentz symmetry breaking $\D_{\rm L\a\b}$ (\ref{lorentzbreak}), can be written as 
\be
\Delta_{{\rm L}\a\b}={W}_{{\rm L}\a\b}\widehat\Delta_{\rm L}~, \label{DeltaL}
\ee
where $\widehat\Delta_{\rm L}$ is an integrated local insertion of UV and IR dimensions 
bounded by $\d\le4$ and $\r\ge4$, respectively. Therefore, $\widehat\Delta_{\rm L}$ can be
reabsorbed in the action as a noninvariant counterterm, order by order, establishing the Lorentz Ward 
identity (\ref{quantum-Lor-Ward}) at the quantum level.

The nilpotency identity ({\ref{nilpot1}) together with
\be
\cs_{\G^{(s-1)}}=\cs_{\widetilde\S^{(s-1)}} + {\co}(\hbar)~,
\ee
implies the following consistency condition for the gauge symmetry breaking $\D_{\rm g}$ (\ref{slavnovbreak}):
\be
\cs_{\widetilde\S^{(s-1)}}\D_{\rm g}=0~,\label{breakcond1}
\ee
and beyond that, $\D_{\rm g}$ also satisfy the constraints:
\ba
&&{\d\D_{\rm g}\over\d b}={\d\D_{\rm g}\over\d\ov c}=\int d^4x \frac{\d\D_{\rm g}}{\d c}= \nonumber\\
&&~~~~~~=W_{\rm rigid}\D_{\rm g}={W}_{\rm L\a\b}\D_{\rm g}=0~.\label{breakcond5}
\ea

The Wess-Zumino consistency condition (\ref{breakcond1}) constitutes a
cohomology problem in the sector of ghost number one.
Its solution can always be written as a sum of a trivial cocycle
$\cs_{\S^{(s-1)}}{\wh\D_{\rm g}}^{(0)}$, where ${\wh\D_{\rm g}}^{(0)}$ has ghost number zero,
and of nontrivial elements belonging to the cohomology of $\cs_{\widetilde\S^{(s-1)}}$
(\ref{slavnovlin}) in the sector of ghost number one:
\be
\D_{\rm g}^{(1)} = {\wh\D_{\rm g}}^{(1)} + \cs_{\widetilde\S^{(s-1)}}{\wh\D_{\rm g}}^{(0)}~.\label{breaksplit}
\ee
However, considering the Slavnov-Taylor operator $\cs_{\widetilde\S^{(s-1)}}$ (\ref{slavnovlin}) and the 
quantum breaking (\ref{slavnovbreak}), it results that $\D_{\rm g}^{(1)}$ exhibits UV and IR dimensions 
bounded by $\d\leq4$ and $\r\geq4$.

From the antighost equation in (\ref{breakcond5}): 
\be
\int d^4x \frac{\d\wh\D_{\rm g}^{(1)}}{\d c}=0~,
\ee
it follows that $\wh\D_{\rm g}^{(1)}$ can be written as 
\be
\wh\D_{\rm g}^{(1)} = \int{d^4 x}~{\cal T}_\m\pa^\m c~, \label{delta-T}
\ee
where ${\cal T}_\mu$ is a rank-$1$ tensor with ghost number zero, with UV and IR dimensions 
bounded by $d\leq3$ and $r\geq3$, respectively. However, the tensor ${\cal T}_\mu$ can be split into two 
pieces:  
\be
{\cal T}_\mu = r_{\rm v} {\cal V}_\mu + r_{\rm p} {\cal P}_\mu~, \label{T-tensor}
\ee
where ${\cal V}_\mu$ is a vector and ${\cal P}_\mu$ is a pseudo-vector, with $r_{\rm v}$ and $r_{\rm p}$ 
being coefficients to be determined. By considering the UV and IR dimensional constraints to be satisfied 
by ${\cal T}_\mu$ (\ref{T-tensor}) together with the conditions upon the Slavnov-Taylor breaking 
$\wh\D_{\rm g}^{(1)}$, given by (\ref{breakcond1}) and (\ref{breakcond5}), it follows that:
\be
{\cal T}_\mu = r_{\rm v} \pa^\r F_{\r\m} + r_{\rm p} \epsilon_{\m\n\r\s}A^\n F^{\r\s} ~. \label{T-tensorf}
\ee
Consequently, substituting (\ref{T-tensorf}) into (\ref{delta-T}), the gauge symmetry breaking $\wh\D_{\rm g}^{(1)}$ reads:
\be
\wh\D_{\rm g}^{(1)} = -\frac{r_{\rm p}}{2} \int{d^4 x}~c\,\epsilon_{\m\n\r\s}F^{\m\n}F^{\r\s} ~, \label{ABBJ}
\ee
which is the well-known (Adler-Bardeen-Bell-Jackiw) ABBJ-anomaly \cite{abbj}. Therefore, up to noninvariant 
counterterms, which are $\cs_{\widetilde\S^{(s-1)}}$-variations of the integrated local insertions 
${\wh\D_{\rm g}}^{(0)}$: 
\be
\D_{\rm g}^{(1)} = \cs_{\widetilde\S^{(s-1)}}{\wh\D_{\rm g}}^{(0)}
-\frac{r_{\rm p}}{2} \int{d^4 x}~c\,\epsilon_{\m\n\r\s}F^{\m\n}F^{\r\s}~.\label{deltaABBJ}
\ee
The anomaly coefficient $r_{\rm p}$ does not get renormalizations \cite{adl-bard,PigSor}, besides that, if it vanishes at the one loop order, it is in fact identically zero, thus it is enough to check its vanishing at that order. However, owing to the fact that the potentially dangerous axial current $j_5^\mu=\ov{\psi}\gamma_{5}\gamma^{\mu}\psi$ is coupled only to the external (classical) field $\b_\mu$ -- and not to any quantum field of the model -- there is no gauge anomaly stemming from \cite{piguet-rouet,Ref11,Ref11'}. Consequently, it follows that the Slavnov-Taylor identity (\ref{quantum-slavnov}) is accomplished at the quantum level.

Concerning the potential anomalies, it can be concluded that the presence of the CPT violating interaction term $\Sigma_{\rm SB}$ (\ref{SB}), which couples the axial fermion current $j_5^\mu=\ov{\psi}\gamma_{5}\gamma^{\mu}\psi$ to a constant vector field $b_\m$, does not induce at any order in perturbation theory, independent of any regularization scheme, neither a Lorentz anomaly nor a gauge anomaly. 

%%%%%%%%%%%%%%%%%%%%%%%%%%%%%%%%%%%%%%%%%%
\subsection{The stability condition: in search for counterterms}
%%%%%%%%%%%%%%%%%%%%%%%%%%%%%%%%%%%%%%%%%%
In order to verify if the action in the tree-approximation ($\widetilde{\Sigma}^{(s-1)}$) is stable 
under radiative corrections, we perturb it by an arbitrary integrated local functional (counterterm) 
$\widetilde\S^{c (s-1)}$, such that
\be
\widehat\S^{(s-1)}=\widetilde\S^{(s-1)}+\ve \widetilde\S^{c (s-1)}~, \label{adef}
\ee
where $\ve$ is an infinitesimal parameter. The functional $\widetilde\S^c\equiv\widetilde\S^{c (s-1)}|_{s=1}$ 
has the same quantum numbers as the action in the tree-approximation at $s=1$.

The deformed action $\widehat\S^{(s-1)}$ must still obey all the conditions presented above, henceforth, 
$\widetilde\S^{c (s-1)}$ is subjected to the following set of constraints:
\ba
&&\cs_{\S^{(s-1)}}\widetilde\S^{c (s-1)}=0~, \label{stabcond}\\
&&{\d\widetilde\S^{c (s-1)}\over{\d b}}={\d\widetilde\S^{c (s-1)}\over{\d{\ov c}}}=
{\d\widetilde\S^{c (s-1)}\over{\d c}}=0~, \label{cond}\\
&&W_{\rm rigid} \widetilde\S^{c (s-1)}=0~, \label{crigidcond}\\
&&{W}_{\rm L\a\b}\widetilde\S^{c (s-1)}=0~. \label{lorentzcond}
\ea

The most general invariant counterterm $\widetilde\S^{c (s-1)}$ -- the most general field polynomial -- with UV and IR dimensions bounded by $\d\le4$ and $\r\ge4$, with ghost number zero and fulfilling the conditions 
displayed in Eqs.(\ref{stabcond})-(\ref{lorentzcond}), reads:
\ba
\widetilde\S^{c (s-1)}\bigg|_{\r\ge 4}^{\d\le 4}&\!\!=\!\!&\int{d^4 x}~\biggl\{\alpha_1 i\ov{\psi}\gamma^{\mu}(\partial_{\mu}+ieA_{\mu})\psi + \nonumber\\
&\!\!+\!\!&\alpha_2 \ov{\psi}\psi + \alpha_3 F^{\mu\nu}F_{\mu\nu} +  \nonumber\\
&\!\!+\!\!&\alpha_4 \left(\beta_{\mu} + b_{\mu}\right)\ov{\psi}\gamma_{5}\gamma^{\mu}\psi \biggr\}~.\label{finalcount}
\ea
The coefficients $\alpha_1,\ldots,\alpha_4$ are arbitrary, and they are fixed, order by order in perturbation 
theory, by the following four normalization conditions:
\ba
\!\!\!\!&&\G_{\ov\psi\psi}({\slashed{p}})\bigg|_{{\slashed{p}}=m}=0~,~~
\frac{\partial}{\partial{\slashed{p}}}\G_{\ov\psi\psi}({\slashed{p}})\bigg|_{{\slashed{p}}=m}=1~,\nonumber\\
&&\frac{\partial}{\partial p^2}\G_{A_\m A_\m}(p^2)\bigg|_{p^2=\kappa^2}^{s=1}=1~,\nonumber\\
&&-\frac14 {\rm Tr}[\gamma^\mu\gamma^5 \G_{\b_\m\ov\psi\psi}(0,{\slashed{p}})]\bigg|_{{\slashed{p}}=m}=1~. \label{normcond}
\ea
Notwithstanding, it shall be stressed here that, a Chern-Simons-like term of the type 
\ba
\S_{\rm CS}\bigg|_{\r\ge 4}^{\d\le 4}&\!\!=\!\!&\int{d^4 x}~\alpha_5\biggl\{\epsilon_{\mu\nu\alpha\beta}\beta^{\mu}A^{\nu}\partial^{\alpha}A^{\beta}\bigg|_4^4 + \nonumber\\
&\!\!+\!\!&\epsilon_{\mu\nu\alpha\beta}b^\mu A^{\nu}\partial^{\alpha}A^{\beta}\bigg|_3^3\biggr\}~, 
\label{CS}
\ea
in spite of fulfils the conditions (\ref{cond})-(\ref{lorentzcond}), its first term breaks gauge invariance by 
violating the Slavnov-Taylor identity (\ref{stabcond}), whereas its second term violates the IR dimension 
constraint ($\r\ge4$), it has IR dimension equal to three. Therefore, the Chern-Simons-like term 
$\S_{\rm CS}$ ($\ref{CS}$) {\it can never be generated by radiative corrections if the renormalization procedure is performed correctly}. First, by taking care of the IR divergences -- for instance, through the 
Lowenstein-Zimmermann method \cite{Low} -- that show up, thanks to the presence of the photon, which is massless. Second, by properly treating and controlling the Lorentz symmetry breaking through the Symanzik method \cite{Sym1,Sym2}. Anyway, even though the external field $\beta_\mu$ was not introduced in order to control the Lorentz breaking, the Chern-Simons-like term -- which is a soft Lorentz breaking (UV dimension less than four) -- would not be radiatively generated as explained above, nevertheless, any gauge invariant hard Lorentz breaking (UV dimension equal to four) could be induced by radiative corrections. In summary, a CPT-odd and Lorentz-violating Chern-Simons-like term {\it is definitely not radiatively induced} at any order in perturbation theory, independent of any regularization scheme, by coupling the axial fermion current $j_5^\mu=\ov{\psi}\gamma_{5}\gamma^{\mu}\psi$ to a constant vector field $b_\m$.

%%%%%%%%%%%%%%%%%%%%%%%%%%%%%%%%%%%%%%%%%%%%%%%%%%%%%%%%%%%%%%%%
\section{Conclusions}
\label{conc}
%%%%%%%%%%%%%%%%%%%%%%%%%%%%%%%%%%%%%%%%%%%%%%%%%%%%%%%%%%%%%%%%
In this work we reassess the discussion on the radiative generation of a Chern-Simons-like term induced from quantum corrections in the extended QED. We prove, to all orders in perturbation theory, that a CPT-odd and Lorentz violating Chern-Simons-like term, {\it definitively}, is not radiatively induced by the axial coupling of the fermions with the constant vector $b_\mu$. The proof of this fact is based on general theorems of perturbative quantum field theory, where the Lowenstein-Zimmermann subtraction scheme in the framework of Bogoliubov-Parasiuk-Hepp-Zimmermann-Lowenstein (BPHZL) renormalization method is adopted.

It is true that we need new ideas to go beyond the Standard Model, for instance, which is the case of the Lorentz symmetry breaking, where if it is manifested or not in our universe has been the subject of much discussion, however so far, no trace was found up to now. Experiments are the final judgement of a theory, which has to be checked experimentally, but the Lorentz symmetry breaking still remains a theoretical construction, regardless of how seductive the idea can be. Nevertheless, even as a theoretical construction, the idea of the Lorentz symmetry breaking should be well grounded and treated properly, although it seems that is not the case in the recent literature on the subject. 

Particularly here, we analyze the issue intensively studied in recent years on the generation of a Lorentz violating Chern-Simons-like term by radiative corrections in the extended QED. 
Unfortunately, several recent works, dealing on the subject, do not consider very carefully the Lorentz symmetry breaking -- neither at the classical level nor at the quantum level -- not taking into account the requirements that Symanzik-Becchi-Rouet-Stora have shown to be necessary. Those authors should read the seminal works by Symanzik-Becchi-Rouet-Stora \cite{Sym1,Sym2,Stora1,Stora2,Stora3} and {\it devour them}.

It shall be stressed that it is urgent and mandatory the reconsideration of the fundamental works on renormalization of quantum field models developed mainly in the 1970's, especially the articles by Symanzik-Becchi-Rouet-Stora on renormalizable models with broken symmetry, 
which provides an appropriate theoretical tool susceptible to avoid some {\it bad} conclusions associated with models with broken Lorentz symmetry. It is important to emphasize that, the main characteristic of this method is the {\it control} of the breaking and, in particular, its power-counting properties, converting the initial action containing terms that violate the Lorentz symmetry into one which is invariant under 
the original transformation by adding external fields (the Symanzik sources). Without this control, the study of the stability (here meant additive renormalization) tells us that {\it any} term that breaks the Lorentz symmetry, compatible with the power-counting, must {\it necessarily} be present in the starting (classical) action. On the other hand, if we include in the initial (classical) action all terms that break the Lorentz symmetry, compatible with the locality and power-counting, no breaking control 
is required (see Ref. \cite{Nos2}). Therefore, paraphrasing Symanzik, {\it whether you like it or not, you have to include in the classical action all Lorentz violating terms consistent with locality and power-counting, unless otherwise constrained by a breaking control}.

%%%%%%%%%%%%%%%%%%%%%%%%%
\section*{Acknowledgments} 
%%%%%%%%%%%%%%%%%%%%%%%%%
The authors dedicate this work to the memory of Prof. Manfred Schweda (1939-2017). O.M.D.C. dedicates this work to his father (Oswaldo Del Cima, {\it in memoriam}), mother (Victoria M. Del Cima, {\it in memoriam}), daughter (Vittoria) and son (Enzo). This work was partially funded by FAPEMIG and CNPq (O.P.).

%%%%%%%%%%%%%%%%%%%%%%%%%%%%%%%%%%%%%%%%%%%%%%%%%%%%%%%%%%%%%%%%%%%%%%%%%%%

%%%%%%%%%%%%%%%%%%%%%%%%%%%%%%%%%%%%%%%%%%%%%%%%%%%%%%%%%%%%%%%%%%%%%%%%%%%%%%

%%%%%%%%%%%%%%%%%%%%%%%%%%%%%%%%%%%%%%%%%%%%%%%%%%%%%%%%%%%%%%%%%%%%%%%%%%%%%%
\end{document}